\documentclass[final,authoryear,3p,times]{elsarticle}
\usepackage{float}
\usepackage{graphicx}

\def\a{true}
\def\useyap{false}
\ifx\a\useyap
\usepackage[dvipdf,
CJKbookmarks=true,
bookmarksnumbered=true,
bookmarksopen=true,
colorlinks=true,
citecolor=blue,
linkcolor=red,
anchorcolor=green,
urlcolor=blue
]{hyperref}
\else
\usepackage[dvipdfm,
CJKbookmarks=true,
bookmarksnumbered=true,
bookmarksopen=true,
colorlinks=true,
citecolor=blue,
linkcolor=red,
anchorcolor=green,
urlcolor=blue
]{hyperref}
\AtBeginDvi{} 
\fi

\usepackage{amssymb}

\journal{\url{arXiv.org}}

\begin{document}

\begin{frontmatter}

\title{The Sensitivity of Respondent-driven Sampling Method}

\author[susocial,kiphs]{Xin Lu\corref{cor1}}
\ead{lu.xin@sociology.su.se}
\author[kiphs]{Linus Bengtsson}
\author[sumath]{Tom Britton}
\author[kimeb]{Martin Camitz}
\author[kim]{Beom Jun Kim}
\author[kiphs]{Anna Thorson}
\author[susocial]{Fredrik Liljeros}

\cortext[cor1]{Address for correspondence: Xin Lu, Department of Sociology, Stockholm University, SE-106 91, Stockholm, Sweden.}

\address[susocial]{Department of Sociology, Stockholm University, Stockholm, Sweden}
\address[kiphs]{Department of Public Health Sciences, Karolinska Institute, Stockholm, Sweden}
\address[sumath]{Department of Mathematics, Stockholm University, Stockholm, Sweden}
\address[kimeb]{Department of Medical Epidemiology and Biostatistics, Karolinska Institute, Stockholm, Sweden}
\address[kim]{Department of Physics, Sungkyunkwan University, Suwon, Korea.}

\begin{abstract}
Researchers in many scientific fields make inferences from individuals to larger groups. For many groups however, there is no list of members from which to take a random sample. Respondent-driven sampling (RDS) is a relatively new sampling methodology that circumvents this difficulty by using the social networks of the groups under study. The RDS method has been shown to provide unbiased estimates of population proportions given certain conditions. The method is now widely used in the study of HIV-related high-risk populations globally. In this paper, we test the RDS methodology by simulating RDS studies on the social networks of a large LGBT web community. The robustness of the RDS method is tested by violating, one by one, the conditions under which the method provides unbiased estimates. Results reveal that the risk of bias is large if networks are directed, or respondents choose to invite persons based on characteristics that are correlated with the study outcomes. If these two problems are absent, the RDS method shows strong resistance to low response rates and certain errors in the participants' reporting of their network sizes. Other issues that might affect the RDS estimates, such as the method for choosing initial participants, the maximum number of recruitments per participant, sampling with or without replacement and variations in network structures, are also simulated and discussed.
\end{abstract}


\end{frontmatter}

\section{Introduction}\label{Sec1 introduction}
Hidden or hard-to-reach populations, such as injecting drug users, men who have sex with men and sex workers, are generally difficult to access because of their strong privacy concerns and a lack of a well-defined sampling frame from which a random sample can be drawn \citep{Heckathorn1997}. Sampling frames are also lacking for many groups without strong privacy concerns, such as for example jazz musicians \citep{Heckathorn2001}. Methods for obtaining information about such groups have involved contacting especially knowledgeable persons within the group (key informant sampling) \citep{Deaux1985}, drawing a sample of participants from locations where group members are known to pass (targeted sampling) \citep{Watters1989}, or asking members of a group to give the researchers contact details of others in the same group (snowball sampling) \citep{Erickson1979}. However, these methods all introduce a considerable selection bias, which impairs generalization of findings from the sample to the studied population \citep{Heckathorn1997,Magnani2005}.

Respondent-driven sampling (RDS) is a method developed to overcome the challenges of selection bias when sampling hidden populations \citep{Heckathorn1997,Heckathorn2002,Salganik2004,Salganik2006,Volz2008}. An RDS study starts out by purposively selecting some participants who are members of the study population (usually 5 to 15). These persons are called ``seeds''. The seeds are given a number of recruitment ``coupons'' (usually 3) to distribute to friends and acquaintances within the study population. If those friends who receive a coupon decide to participate, they are in turn given the same number of coupons to invite further participants. Participants are rewarded for their personal participation in the study as well as for each peer they invite and who also participates. The invitation coupon contains a serial number that enables the researchers to follow the recruitment chains in the sample. If the recruitment chains are sufficiently long, the sample composition stabilizes and becomes independent of the characteristics of the seeds. What' s more, each participant is asked for the number of persons he or she knows within the study population, known as his/her ``personal network size'' or ``degree''. The degree of a participant is important to collect as participants with large degrees are oversampled and participants with small degrees are undersampled. Knowing the degree of each participant thus allows adjustment for this bias.

When the sample has been collected, the proportion of persons with the characteristic $A$ in the population can be estimated by the \textit{RDS\uppercase\expandafter{\romannumeral2}} estimator \citep{Volz2008}:

\begin{equation}\label{eq1}
\hat P_A  = {{\sum\limits_{i \in A \cap S} {d_i^{ - 1} } } \mathord{\left/
 {\vphantom {{\sum\limits_{i \in A \cap S} {d_i^{ - 1} } } {\sum\limits_{i \in S} {d_i^{ - 1} } }}} \right.
 \kern-\nulldelimiterspace} {\sum\limits_{i \in S} {d_i^{ - 1} } }}
\end{equation}
		
Where $d_i$ is the degree of individual $i$, and $S$ the set of sampled individuals.

\cite{Volz2008} proved that the \textit{RDS\uppercase\expandafter{\romannumeral2}} estimator provides asymptotically unbiased estimates if the following assumptions are fulfilled:

\begin{enumerate}[(i)]
\item Reciprocity: individuals in the studied population maintain and recruit peers through reciprocal relationships, that is, the network within which recruitment happens, is undirected;
\item Connectedness: each individual in the studied population has a chance to be invited to participate, that is, the network forms a single component;
\item Sampling is with replacement: individuals are allowed to be recruited into the sample more than once;
\item Degree: respondents can accurately report their degree in the network;
\item Random recruitment: peer-recruitment is a random selection from the respondents' personal network;
\item Each respondent recruits a single peer, that is, the number of recruitment coupons is one.
\end{enumerate}

The ability of making generalizable estimates together with a feasible field implementation have contributed to a rapid increase in RDS studies conducted globally in recent years: to date, well over 100 studies in over 30 countries have been performed \citep{Johnston2008, Malekinejad2008,Wejnert2008}.

However, the assumptions underlying the RDS estimator are not easily met in real life. First, most social networks contain directed edges, or edges which do not have the same strength in each direction. Second, to prevent participants from colluding to recruit each other back and forth to gain rewards, empirical RDS applications sample without replacement, meaning that respondents can only participate once. Third, it is difficult for respondents to report their degree accurately. Fourth, participants usually pass their coupons to peers with whom they have a close rather than a more distant relationship, which is not a random selection. Fifth, to avoid recruitment chains stopping too early, researchers most often use three coupons rather than one.

While rationality of those assumptions have been questioned and argued in the literature \citep{Heckathorn1997,Heckathorn2002,Salganik2004,Salganik2006,Volz2008,Heimer2005,Goel2009}, it is difficult to assess the reliability of RDS since the study population is usually unknown. Previous studies have mainly used artificially constructed networks and have often been linked to the introduction of new estimators, hence they have focused on different aspects and did not cover the whole scope of possible violations. The most recent and comprehensive study was made by \cite{Gile2009}. Using artificially networks, which were constructed from pilot data from the CDC surveillance program \citep{Abdul2006}, they simulated RDS with respect to the bias induced by the violations of assumptions \romannumeral3, \romannumeral5\ and \romannumeral6. The population size they used was quite small (1000) and the fractions of sample sizes are relatively large, from 50\% to 95\% (500 up to 950). They addressed the possibility of reduction of bias by discarding early waves and found a potential bias caused by preferential selection of peers and sampling without replacement. However, the number of seeds, waves and coupons were fixed, and they didn't discuss other assumptions that might affect the RDS estimates, such as directedness of networks, recruitment failure, and degree reporting error amongst others.

Based on the increasing use of RDS within research, together with plausible real life difficulties in completely fulfilling the theoretical assumptions, we identified a need for systematically testing the robustness of the RDS method when sampling diverges from the basic assumptions in the analytical proof. In this study, we simulate RDS studies within a real-life social network, which is constructed by data extracted from a lesbian, gay, bisexual, transgender (LGBT) web community. By violating the RDS assumptions one by one, we seek to analyze the resistance of the estimator to bias and to evaluate how real life deviations from theoretical requirements will affect the estimates. We use the \textit{RDS\uppercase\expandafter{\romannumeral2}} estimator for all RDS estimates in this article as this estimator has improved analytical powers compared to earlier RDS estimators and provides equivalent estimates when data-smoothing is used \citep{Heckathorn2007,Volz2008,Gile2009}.

Four measurements are used throughout this paper: the average estimate (AE), defined by the mean of the \textit{RDS\uppercase\expandafter{\romannumeral2}} estimates,
$AE_j  = \sum\nolimits_{i = 1}^m {est_{ij} /m} $
, where $est_{ij}$ is the estimate of \textit{RDS\uppercase\expandafter{\romannumeral2}} at the $i^{th}$ simulation when sample size is $j$; the bias, defined by the absolute difference between AE and the true population, $Bias_j  = |AE_j  - P^* |$; the standard deviation (SD) of estimates for a given sample size $j$, $SD_j$; and finally the mean absolute error (MAE) of estimates for a given sample size $j$,
$MAE_j  = \sum\nolimits_{i = 1}^m {|est_{ij}  - P^* |/m} $.

The rest of this paper is organized as follows: in section \ref{The MSM Network}, we give a brief description of our data and networks; in section \ref{RDS on the Undirected Network}, we describe the results of simulating RDS studies in the networks when all of the assumptions are satisfied; in section \ref{Violations of Assumptions}, we test the effects of violating the assumptions one by one; and in section \ref{Conclusions}, we summarize and draw our conclusions.

\section{The MSM Network}\label{The MSM Network}

\textit{Data collection}. ``Qruiser''(\url{www.qx.se}), is the Nordic region's largest and most active web community for homosexual, bisexual, transgender, and queer persons. Contacts between members on the web site are maintained mainly by a ``favorite list'', on which each member can add any other member without approval from that member. Members can attend clubs (web pages about specific topics) and send messages to each other \citep{Rybski2009}.

We collected information on personal profiles as well as on all messages sent within the web-community from Dec 15, 2005, to Jan 18, 2006. During the 63 days of this data collection period, 12,590,911 messages were recorded and there were 184,819 distinct members registered on the web site.

\textit{Network Formation}. Based on the membership profiles, we extracted a network that contained only members characterizing themselves as homosexual males. We define an outgoing edge to be formed if a member has another member on his favorite list. An edge is called reciprocal if a connected pair of members have both an ingoing and an outgoing edge between each other. If a pair does not have both an ingoing and an outgoing edge between each other, it is called irreciprocal. To avoid the inclusion of inactive persons, members were required to have sent at least one message to any other person on the site during the data collection period.

\begin{figure}[htb]
\centering
\includegraphics[width=1\textwidth]{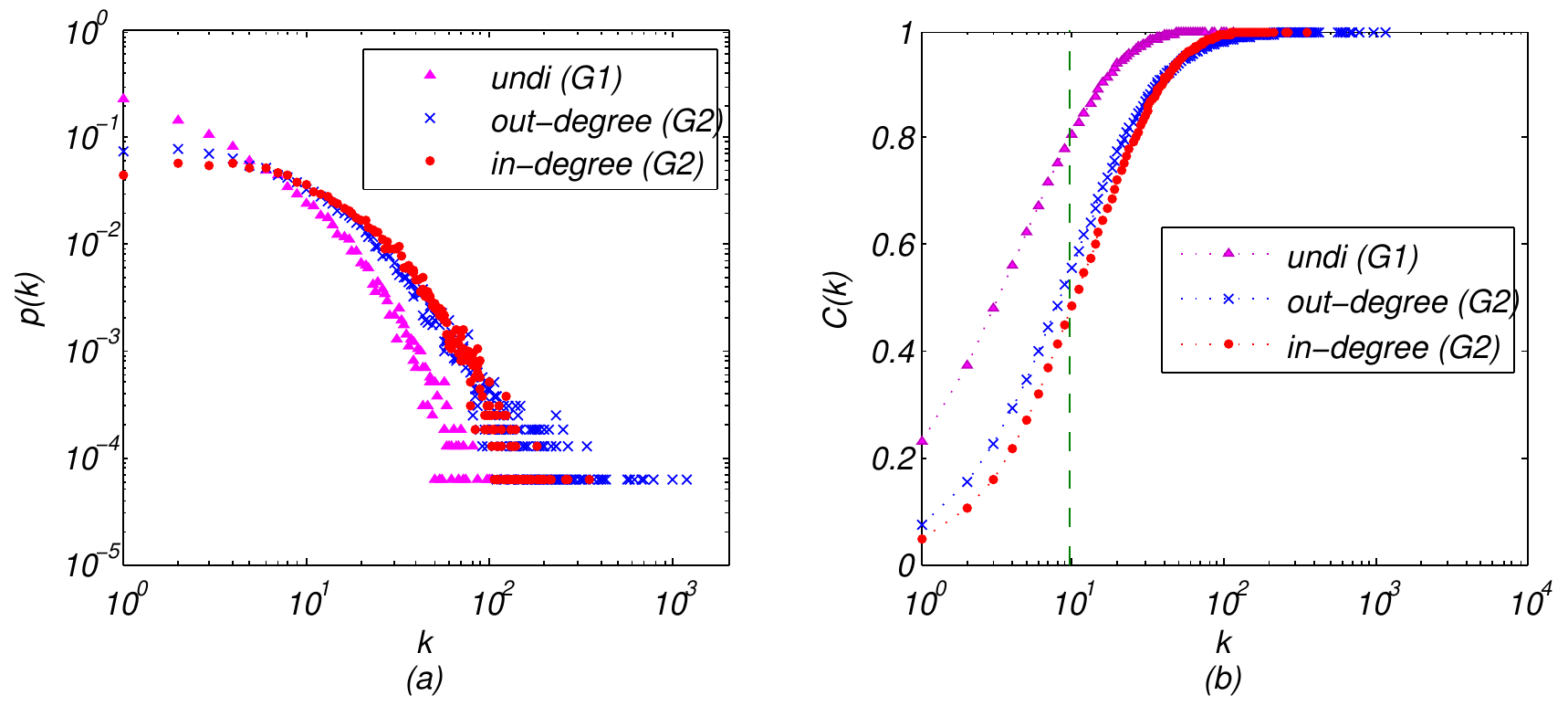}
\caption{\label{fig1}(a) Degree distribution (b) Cumulative degree distribution}
\end{figure}

For our research purpose, only members of the Giant Connected Component (GCC), defined by the largest component connected with only reciprocal edges, were kept as nodes (16,082 nodes). If we keep only the reciprocal edges in that GCC, we obtain an undirected network ($G1$) with an average degree of 6.74. If we keep both reciprocal and irreciprocal edges, we obtain a directed network ($G2$) with an average degree of 17.2. Note that the definition of the GCC ensures that all nodes have a chance to be recruited with RDS sampling in both $G1$ and $G2$. Degree distributions for both $G1$ and $G2$ are plotted in \autoref{fig1}. The distributions are very skewed, e.g., half of the members in $G2$ have no more than 10 outgoing edges, while a small proportion of members have a large number of outgoing edges.

\textit{Homophily}. An important issue for chain referral sampling is the homophily of edge formation, which is the probability for participants of connecting with friends that are similar to themselves rather than connecting randomly \citep{Rapoport1980,Morris1995, McPherson2001,Heckathorn2002}. The homophily of different groups in our network are shown in \autoref{table 1}. The homophily with respect to \textit{age} and \textit{county} are fairly large, indicating a fair part of the edges being formed between members of the same \textit{age} or between members living in the same \textit{county}. Taking \textit{county} within the undirected network, $G1$ as an example, members who live in Stockholm formed edges with members who also lived in Stockholm 50\% of the time, while they formed edges randomly among all cities (including Stockholm) the remaining 50\% of the time. The homophily for persons living Stockholm is thus 0.5. Homophilies for \textit{civil status} and \textit{profession} were very small, indicating that edges were formed as if members, regarding \textit{civil status} and \textit{profession}, chose randomly among other members.

\begin{table}
\setlength{\abovecaptionskip}{0pt}
\setlength{\belowcaptionskip}{10pt}
\caption{\label{table 1}Proportions ($P^*$) and homophilies ($H$) of groups in the undirected ($G1$) and directed ($G2$) networks}
\centering
\begin{tabular}{ccccccccccccc}
\hline
\ &\ &\multicolumn{2}{c}{\bf Age}&\ &\multicolumn{2}{c}{\bf County}&\ &\multicolumn{2}{c}{\bf Civil Status}&\ &\multicolumn{2}{c}{\textbf{Profession}}\\
\cline{3-13}
\ &\ &Before 1980&others&\ &Stockholm&others&\ &Single&others&\ &Employed&others\\
\hline
$P^*$&\ &77.77\%&22.23\%&\ &38.79\%&61.21\%&\ &40.39\%&59.61\%&\ &38.19\%&61.81\%\\
$H G1$&\ &0.40&0.37&\ &0.50&0.40&\ &0.05&0.08&\ &0.13&-0.05\\
$H G2$&\ &0.23&0.34&\ &0.50&0.28&\ &0.03&0.07&\ &0.06&0.02\\
\hline
\end{tabular}
\end{table}

\textit{Network Variation}. To avoid misleading conclusions resulting from the effects of network structure and edge density in our simulations on the undirected network ($G1$), we created two variants of $G1$: the first type of networks ($G1_{add}$) was obtained by randomly adding reciprocal edges with properties proportional to $G1$ until the average degree was increased by 20, for each property, separately. The second type of networks ($G1_{rand}$) was obtained by randomly rewiring each pair of reciprocal edges to another node with the same property as the former one. After the procedures above, we obtained four denser networks and four randomized networks, all with the homophily unchanged for each property, respectively. The changes in degree distributions for both $G1_{add}$ and $G1_{rand}$ can be found in \hyperlink{App-A}{Appendix A}.

Moreover, to test the effects of preferential recruitment in RDS (section \ref{Preferential recruitment}), we weighted each reciprocal edge in $G1$ in two ways: by the maximum number of sent messages in any one direction, and by the minimum number of sent messages in any one direction. For example, if node $A$ sent 10 messages to node $B$ and received 5 back from $B$, the weight on edge $e_{A,B}$ ($e_{B,A}$) would be 10 for the maximum-weighted network ($G1_{max}$) and 5 for the minimum-weighted network ($G1_{min}$). In these two weighted networks, respondents were supposed to recruit peers with probability proportional to the edge weights.

We now proceed to describe the results of simulated RDS samplings under variable circumstances in  the networks described above. We compare the true population proportions of the four variables in \autoref{table 1} (two with high homophily and two with low) with the RDS estimates given by the simulated samplings. All simulations are repeated 10,000 times unless otherwise stated.

\section{RDS on the Undirected Network}\label{RDS on the Undirected Network}

\begin{figure}[htb]
\centering
\includegraphics[width=1\textwidth]{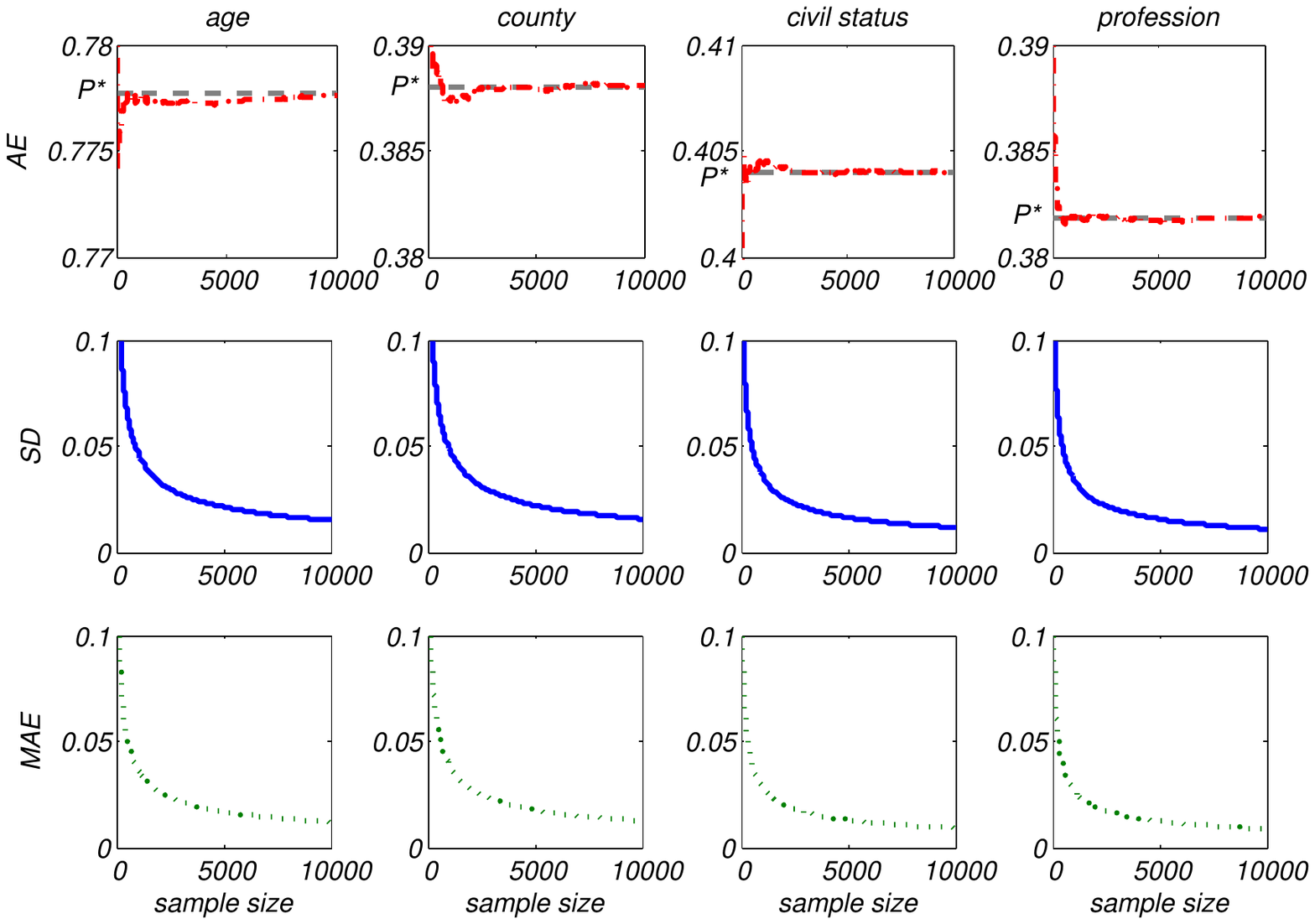}
\caption{\textit{RDS\uppercase\expandafter{\romannumeral2}} estimations on the undirected network ($G1$). The average estimates approached the true proportions very fast. When the sample size was 500 the Bias was only 0.0002, 0.0009, 0.00002, 0.0002 for \textit{age}, \textit{county}, \textit{civil status} and \textit{profession}, respectively.}\label{fig2}
\end{figure}

We first ran simulations on the undirected network, $G1$, to see whether the \textit{RDS\uppercase\expandafter{\romannumeral2}} estimator worked well when all the earlier stated assumptions (\romannumeral1-\romannumeral6) were satisfied. We started each simulation with a single randomly selected seed and we restricted the number of coupons to one, that is, each recruiter could only recruit one other person (assumption \romannumeral6). All respondents were selected randomly from the recruiters' personal network (assumption \romannumeral5), and nodes could be selected multiple times (sampling with replacement, assumption \romannumeral3). Since all participants' degrees were assumed to be known by the participants themselves, and $G1$ is a single connected component with only reciprocal edges, assumptions \romannumeral1, \romannumeral2\ and \romannumeral4\ were also satisfied. We kept recruiting participants in the simulation until the sample size reached 10,000 participants. The average estimation, standard deviation, and mean absolute error are shown in \autoref{fig2}. Even though our network is sparse, the \textit{RDS\uppercase\expandafter{\romannumeral2}} estimates converged to the true population proportions ($P^*$) very quickly. The bias in the average estimate for \textit{age}, \textit{county}, \textit{civil status}, and \textit{profession} is shown below in \autoref{fig2}. \hyperlink{App-B}{Appendix B} shows clearer figures for sample sizes of less than 1000.

The SD and MAE decreased to 2\% when the sample size approached 10,000. However, it is rarely possible to recruit this many respondents in a real life RDS study. Reported sample sizes of real life RDS studies are virtually all less than 1000. For our network, we can see that the SD was around 5\%, and the MAE was around 4\% when the sample sizes were between 500 and 1000 participants.

We should note that our network is far from ideal: first, it is sparse compared to reported studies since the average degree is low, only 6.74; second, the degree distribution is skewed with almost 40\% of network members having a degree no higher than 2; third, there is a high homophily with regard to the variables \textit{age} and \textit{county}. Despite these difficulties, the high precision and rapid convergence of the RDS estimates in Figure 2 reveal that the RDS estimator is asymptotically unbiased and works well on undirected networks.

\section{Violations of Assumptions}
\label{Violations of Assumptions}

\subsection{RDS on directed networks}
\label{RDS on directed networks}

If a directed network forms a Giant Strongly Connected Component (GSCC) \citep{Schwarte2002} in which every node can be reached by any other, and assumptions \romannumeral3\ to \romannumeral4\ are satisfied, we can model the RDS sampling as a Markov process, which has the equilibrium:

\begin{equation}\label{eq2}
[x_1 ,x_2 , \ldots ,x_N ] \times \left[ {\begin{array}{*{20}c}
   0 & {e_{12} /d_1^o } &  \cdots  & {e_{1N} /d_1^o }  \\
   {e_{21} /d_2^o } & 0 &  \cdots  & {e_{2N} /d_2^o }  \\
    \vdots  &  \vdots  &  \ddots  &  \vdots   \\
   {e_{N1} /d_N^o } & {e_{N2} /d_N^o } &  \cdots  & 0  \\

 \end{array} } \right] = [x_1 ,x_2 , \ldots ,x_N ]
\end{equation}

Abbreviated as:
\begin{equation}\label{eq3}
X^T  \times A = X^T
\end{equation}

Where $X$ is the equilibrium of the Markov Process, $e_{ij}=1$ if there is an edge from $i$ to $j$, otherwise $e_{ij}=0$, and $d_i^o$ is the out-degree of $i$.

It can be easily verified that

\begin{equation}\label{eq4}
x_i  = d_i /\sum\limits_{j = 1}^N {d_j }
\end{equation}
		
is the solution for Eq.~(\ref{eq3}) if the out-degree and in-degree are equal for all nodes. Actually, Eq.~(\ref{eq4}) is the underlying equation from which the \textit{RDS\uppercase\expandafter{\romannumeral2}} estimator is derived \citep{Volz2008,Goel2009}.

However, the equilibrium can hardly be constructed for general directed networks. We can rewrite Eq.~(\ref{eq3})as:

\begin{equation}\label{eq5}
A^T  \times X = X
\end{equation}

This means that $X$ should be an eigenvector of eigenvalue 1 for $A^T$. (The existence of eigenvalue 1 for $A^T$ can be easily proved as the all-one vector is the eigenvector for $A$ \citep{Woess1994,Brin1998,Page1999}.

Let $V = [v_1 ,v_2 , \ldots ,v_N ]$ be the normalized eigenvector of $A^T$ for eigenvalue 1, then an RDS sample
$\{ s_1 ,s_2 , \ldots ,s_n \} $, can be weighted by

\begin{equation}\label{eq6}
\hat P_A  = \frac{{\sum\limits_{s_i  \in A} {\frac{1}
{{v_{s_i } }}} }}
{{\sum\limits_{s_j } {\frac{1}
{{v_{s_j } }}} }}
\end{equation}
		
to estimate the proportion of individuals in group $A$ in the population. We denote Eq.~(\ref{eq6}) as `\textit{eig}' to weight the RDS samples in a directed network. Note that we can hardly know the value $V$ from an RDS sample.

\begin{figure}[htb]
\centering
\includegraphics[width=1\textwidth]{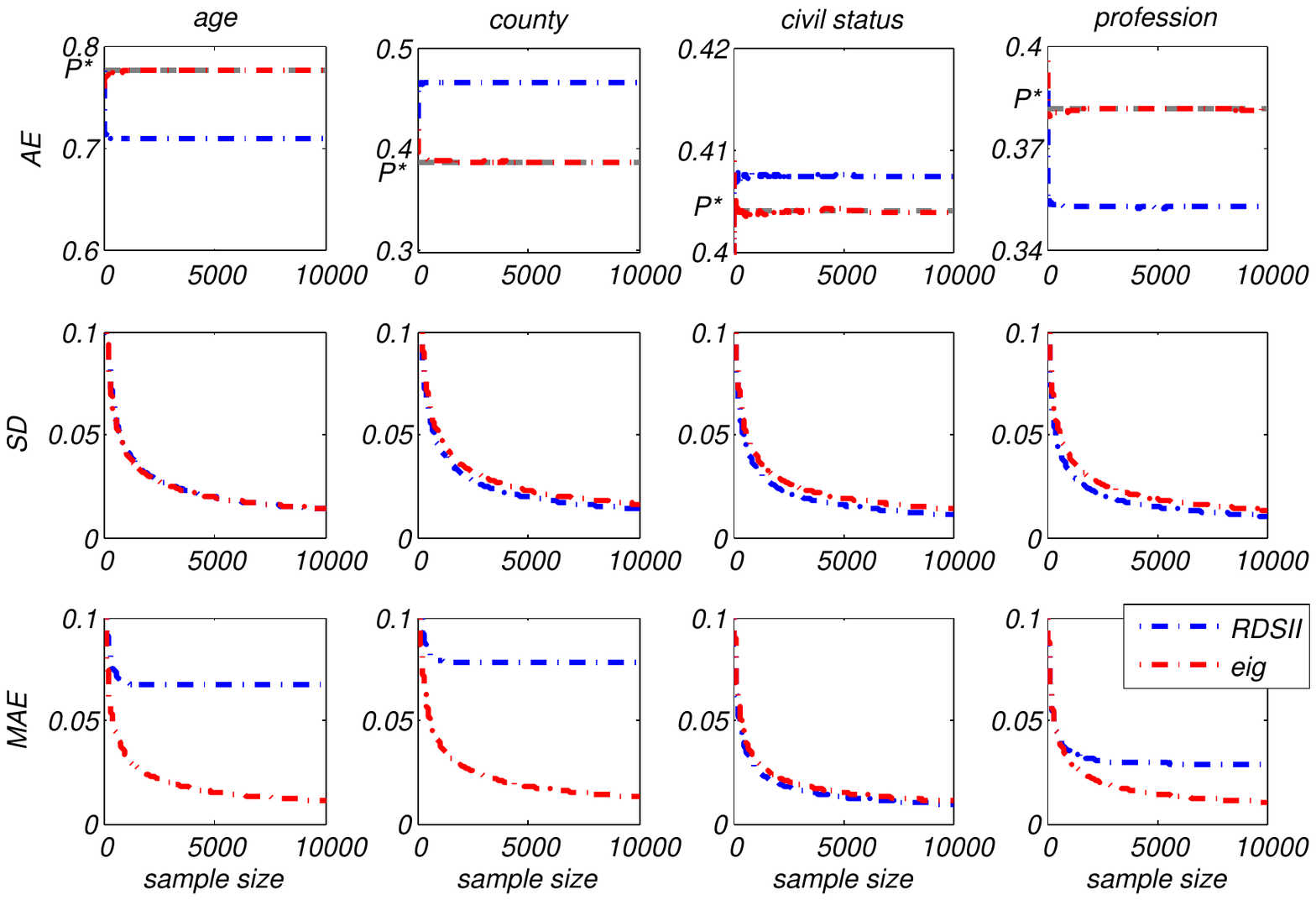}
\caption{Estimations on the directed network ($G2$). Number of seeds=1, coupons=1, with replacement. Blue lines stand for the \textit{RDS\uppercase\expandafter{\romannumeral2}} estimates and red lines for estimates weighted by eigenvectors.}\label{fig3}
\end{figure}

Both \textit{RDS\uppercase\expandafter{\romannumeral2}} and \textit{eig} estimations on the directed network ($G2$) are presented in \autoref{fig3}. Not surprisingly, the \textit{RDS\uppercase\expandafter{\romannumeral2}} estimates were biased for all groups. For \textit{age} and \textit{county}, these biases were as high as 6\%, while \textit{civil status} and \textit{profession} performed better at 0.5\%, and 2.2\%, respectively. However, the \textit{eig} estimates weighted by Eq.~(\ref{eq6}) agreed well with the true population proportions.

The standard deviations (SD) were similar for all four groups (and very similar to the SD of the undirected networks). However, the MAE of \textit{RDS\uppercase\expandafter{\romannumeral2}} in the directed network was much higher than that of the undirected networks for \textit{age} and \textit{county}, ~7\%-8\%, indicating that if the network under study is partly directed, the use of \textit{RDS\uppercase\expandafter{\romannumeral2}} estimations could result in relatively large errors. We can see that the difference for \textit{civil status} and \textit{profession} were very small, telling us that directedness of edges probably has little effect on \textit{RDS\uppercase\expandafter{\romannumeral2}} for groups with low homophily.

\subsection{Sampling without replacement}
\label{Sampling without replacement}

\begin{figure}[htb]
\centering
\includegraphics[width=1\textwidth]{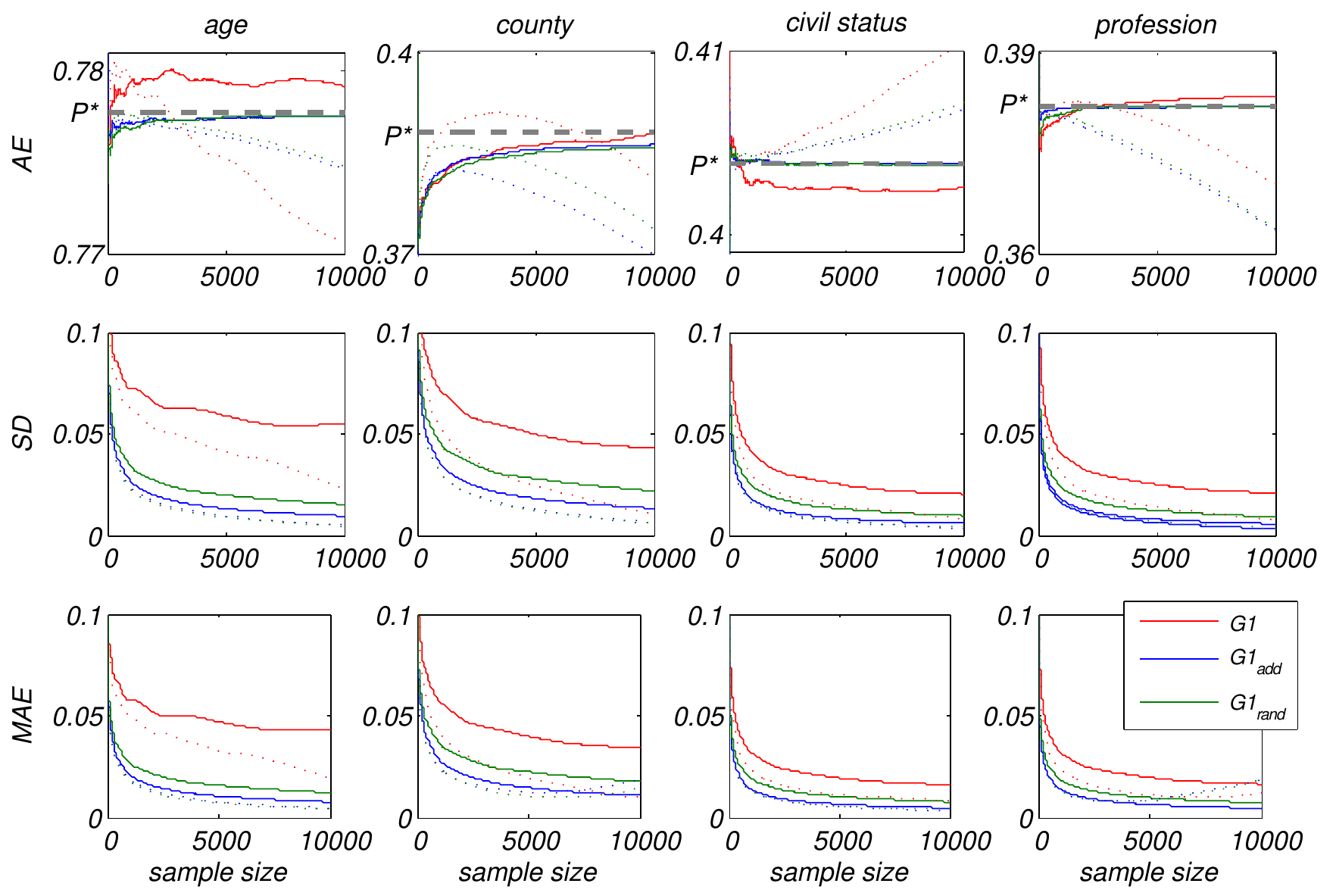}
\caption{Effects of network structures and replacement. Number of seeds=10, coupons=3. Seeds were randomly selected at the beginning of each simulation. Solid lines represent sampling with replacement and dashed lines represent sampling without replacement.}\label{fig4}
\end{figure}

It is generally believed that sampling without replacement (SWOR) creates negligible bias compared to sampling with replacement (SWR) in RDS when the sample size is small relative to the population \citep{Heckathorn1997,Heckathorn2002,Volz2008}. We tested this proposition in our undirected network. What's more, to increase the generalizability of our results, we also compared the replacement effect on $G1_{add}$ and $G1_{rand}$. Results are shown in \autoref{fig4}. We can see from the figures that the \textit{RDS\uppercase\expandafter{\romannumeral2}} estimations for SWR were asymptotically unbiased on all networks, while the estimations for SWOR were biased in different directions in the different networks.

The hypothesis above seemed to hold when sample sizes were small, that is, between 500 and 1000 (see enlarged figure in \hyperlink{App-A}{Appendix C}. The maximum differences in the average estimates between SWR and SWOR were all within 1\%. For \textit{county}, \textit{civil status} and \textit{profession}, the average estimates of SWOR are even slightly closer to the true population than SWR. The SWOR always has a smaller SD and MAE than SWR and this is especially apparent in $G1$. Simulations not included in this paper indicate that networks with skewed degree distribution result in larger variances than those with a Poisson distribution and as networks gets denser, the variance becomes smaller.

\subsection{Rejecting invitations and forgetting peers}
\label{Rejecting invitations and forgetting peers}

While a great majority of participants in RDS studies report a social network size larger than three, far from all distributed coupons result in study participations \citep{Johnston2008,Malekinejad2008}. This could be seen as a violation of assumption \romannumeral4, which states that participants can accurately report their personal network size (assumed to reflect chances of being invited or of the number or peers who have a chance to be invited by the person) and assumption \romannumeral6, which states that all participants use their one coupon to make one successful recruitment. We can note that the latter assumption includes the dual assumption that each coupon generates one further participant and also that each participant receives only one coupon. In our simulations however, three coupons were used, see below. We modeled the effects of deviating from these ideal assumptions by letting each invited member have a probability of rejecting invitations. For each invited member, the probability of rejecting an invitation is called $p_r$. A rejected coupon was discarded and not reused for recruiting a new member.

In addition recruiters could potentially have difficulties remembering all of their friends when considering whom to invite and when estimating the sizes of their personal networks. We modeled this by letting the recruiters ignore some of his/her edges when inviting members from his/her personal network. In the simulations, each edge of a recruiter was given a probability of being ignored. We called this probability $p_i$. An ignored edge had a zero probability of being selected and was not included in the network size of the participant when calculating the \textit{RDS\uppercase\expandafter{\romannumeral2}} estimates.

\begin{figure}[htb]
\centering
\includegraphics[width=1\textwidth]{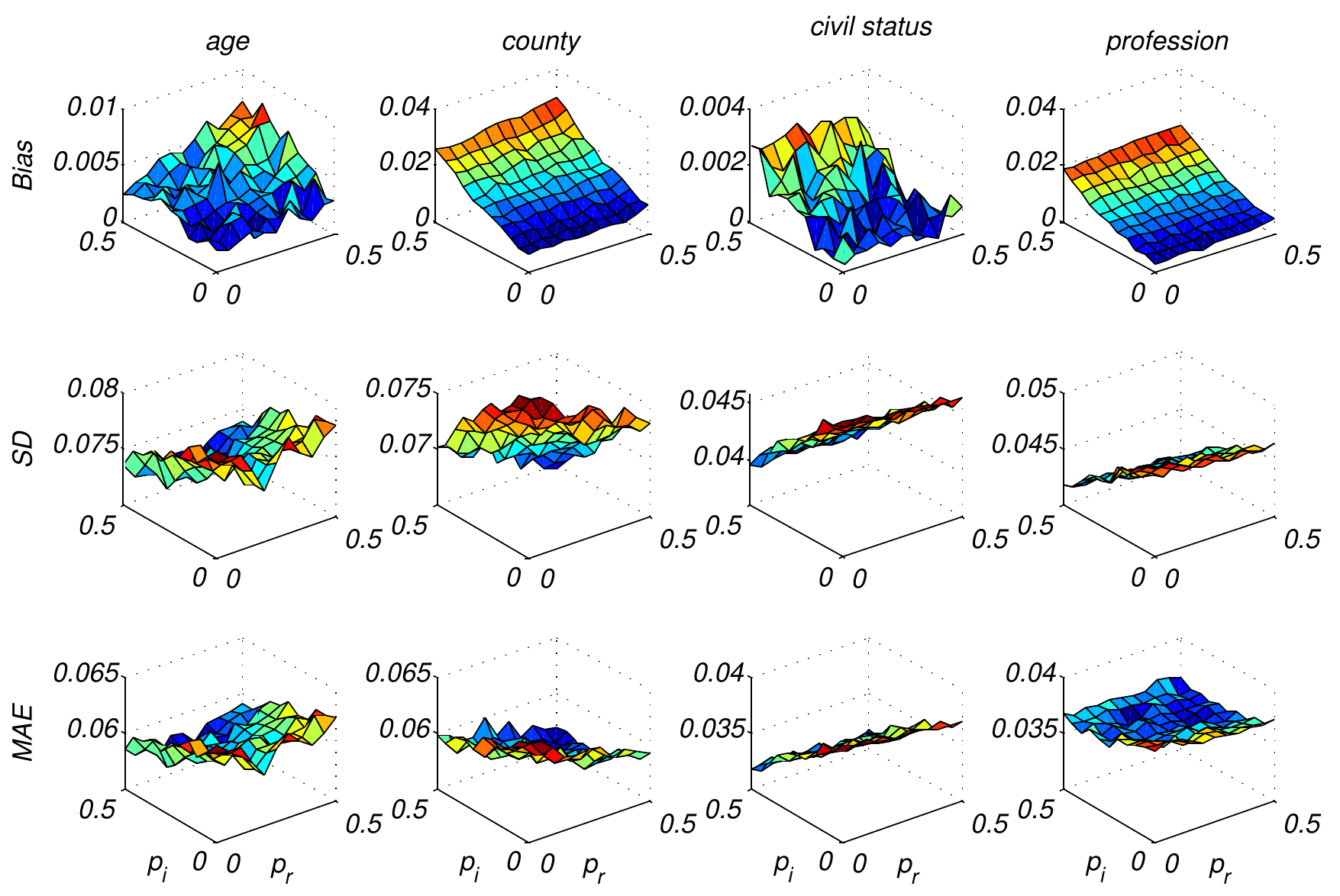}
\caption{Sampling with ignore and reject probabilities in the original, undirected network ($G1$). Simulation was repeated 10,000 times for each combination, number of seeds=10, coupon=3, with replacement. Seeds were randomly selected at the beginning of each simulation. Sample size is 500.}\label{fig5}
\end{figure}

Additionally, we set the number of seeds to 10 and coupons to 3 to make sure that the process could recruit a sample when $p_r$ and $p_i$ became large. Simulation results for the $G1$, $G1_{add}$, and $G1_{rand}$ are displayed as surfaces in \autoref{fig5} and in \hyperlink{App-D1}{Appendix D1}.

When recruitment takes place with ``rejecting'' and ``ignoring'' in the original sparse network (\autoref{fig5}), and in the $G1_{rand}$ networks (\hyperlink{App-D1}{Appendix D1}), which are also sparse, the bias is small to moderate (up to 0.03). Simulation results on the edge-added networks reveal that in these networks, changes in $p_r$ and $p_i$ do not affect the bias \hyperlink{App-D1}{Appendix D1}. Effects on MAE are small in all networks (below 0.01).

In \autoref{fig5} it is also interesting to observe that while increases in $p_r$ and $p_i$ increased the bias, the MAE actually decreased. The small differences seen when varying $p_r$ and $p_i$, as well as the observed decreases in SD and MAE with increasing $p_r$ and $p_i$, all imply a strong resistance of \textit{RDS\uppercase\expandafter{\romannumeral2}} against these recruitment errors, as long as the recruitment chains are able to continue and generate the target sample size.

It should be noted that these simulations do not test all types of violations of assumption \romannumeral4\ and \romannumeral6. If we for example set $p_i$ and $p_r$ as dependent on any of our four outcome variables, errors could be much larger than described above. To evaluate these effects, we performed further simulations in which both the ignore and the reject probabilities differed, depending on group membership.

Let $p_i$ be the probability that a member in the group of interest will be ignored by his friends when these friends are given the possibility to recruit, and let $p_r$ be the probability that a coupon will be rejected by a member in the group of interest. Similarly, let $p'_i$ and $p'_r $ and  be the corresponding ignore and reject probabilities for members in groups of noninterest. Surfaces for RDS with unequal recruiting probabilities are presented in \hyperlink{App-D2}{Appendix D2}. We can see that when the ignore or reject probabilities depended on the characteristics of the members, the RDS estimates gave large bias and error. Take the first simulation in \hyperlink{App-D2}{Appendix D2}, for example: when members born before 1980 rejected half of the invitations given to them and the members born after 1980 did not reject any invitations ($p_i$,and $p'_i $ both set to 0), the bias was over 0.3 for \textit{age}.

When $p_i  = p'_i $, the Bias and MAE are small as long as $p_r  = p'_r $ and vice versa. As both the ignore and reject actions will reduce the inclusion probabilities of group members, they have similar effects on the RDS estimates and can compensate for each other. For example, when the fixed ignore probabilities were $p_i=0.1$, $p'_i=0.3$, the minimum Bias and MAE were when $p_r  > p'_r $. For all the simulations, the values of MAE are almost the same as those of the Bias, revealing that when the studied groups hold different ignore or reject probabilities, the RDS estimates will virtually always be too high or too low. Although differences between groups in $p_i$ can be compensated for by inverse differences between the groups in $p_r$, it can be hypothesized that such a combination of probabilities would be unusual in real life. As participants in RDS studies are rewarded for successful recruitments, a rational and self-interested participant would seek to ignore contacts whom he/she considers unlikely to accept an invitation. This would mean that groups with high $p_r$ would also have a high $p_i$. Unfortunately, such combinations of $p_i$ and $p_r$ always give rise to the largest bias and MAE.

\subsection{Preferential recruitment}
\label{Preferential recruitment}

According to Eq.~(\ref{eq2}), it is easy to infer that the \textit{RDS\uppercase\expandafter{\romannumeral2}} estimator would be biased when recruitment is a non-random selection among the edges of each node, as Eq.~(\ref{eq4}) is no longer the equilibrium.

\begin{figure}[htb]
\centering
\includegraphics[width=1\textwidth]{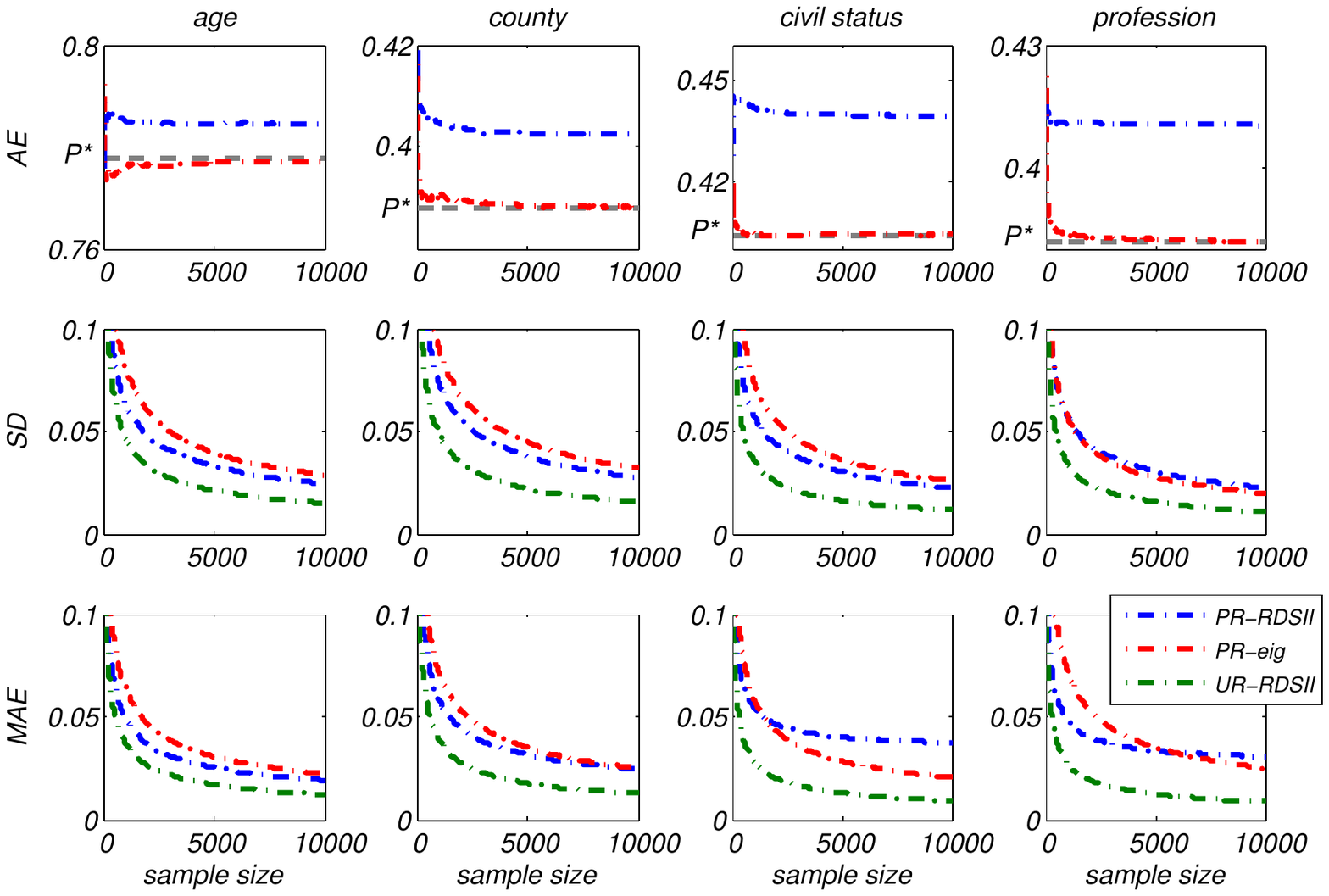}
\caption{RDS on $G1_{max}$ with preferential recruitment. Number of seeds=1, coupons=1, sampling with replacement. Seeds were randomly selected at the beginning of each simulation. The blue lines represent estimations by \textit{RDS\uppercase\expandafter{\romannumeral2}} and red lines represent estimations by eigenvector. Green lines are the \textit{RDS\uppercase\expandafter{\romannumeral2}} estimations for recruitment with uniform probability. Dashed gray lines indicate the true population values.}\label{fig6}
\end{figure}

A plausible non-random recruitment scenario would be that contacts with whom the recruiter interacts more frequently have a higher probability of being invited than those with whom the recruiter interacts only seldom. Simulation results for RDS on $G1_{max}$, in which respondents were supposed to recruit peers with a probability proportional to the edge weights, are presented in \autoref{fig6}. We can see that the \textit{RDS\uppercase\expandafter{\romannumeral2}} estimations were no longer unbiased for all groups. The biases were around 1\%, 2\%, 4\%, and 3\% for the four groups, respectively. When we compare the SD and MAE of preferential recruitment (PR) with uniform recruitment (UR), the preferential recruitment had larger SD and MAE values for all groups, indicating that if the respondents prefer to distribute the coupons to their closer friends, larger \textit{RDS\uppercase\expandafter{\romannumeral2}} estimation errors might occur. Again when we use the \textit{eig} estimations on the weighted network, they agree well with the true population.

Simulations on $G1_{min}$ reveal similar outcomes, see \hyperlink{App-E}{Appendix E}.

\subsection{Effect of seeds and coupons}
\label{Effect of seeds and coupons}

\begin{figure}[htb]
\centering
\includegraphics[width=1\textwidth]{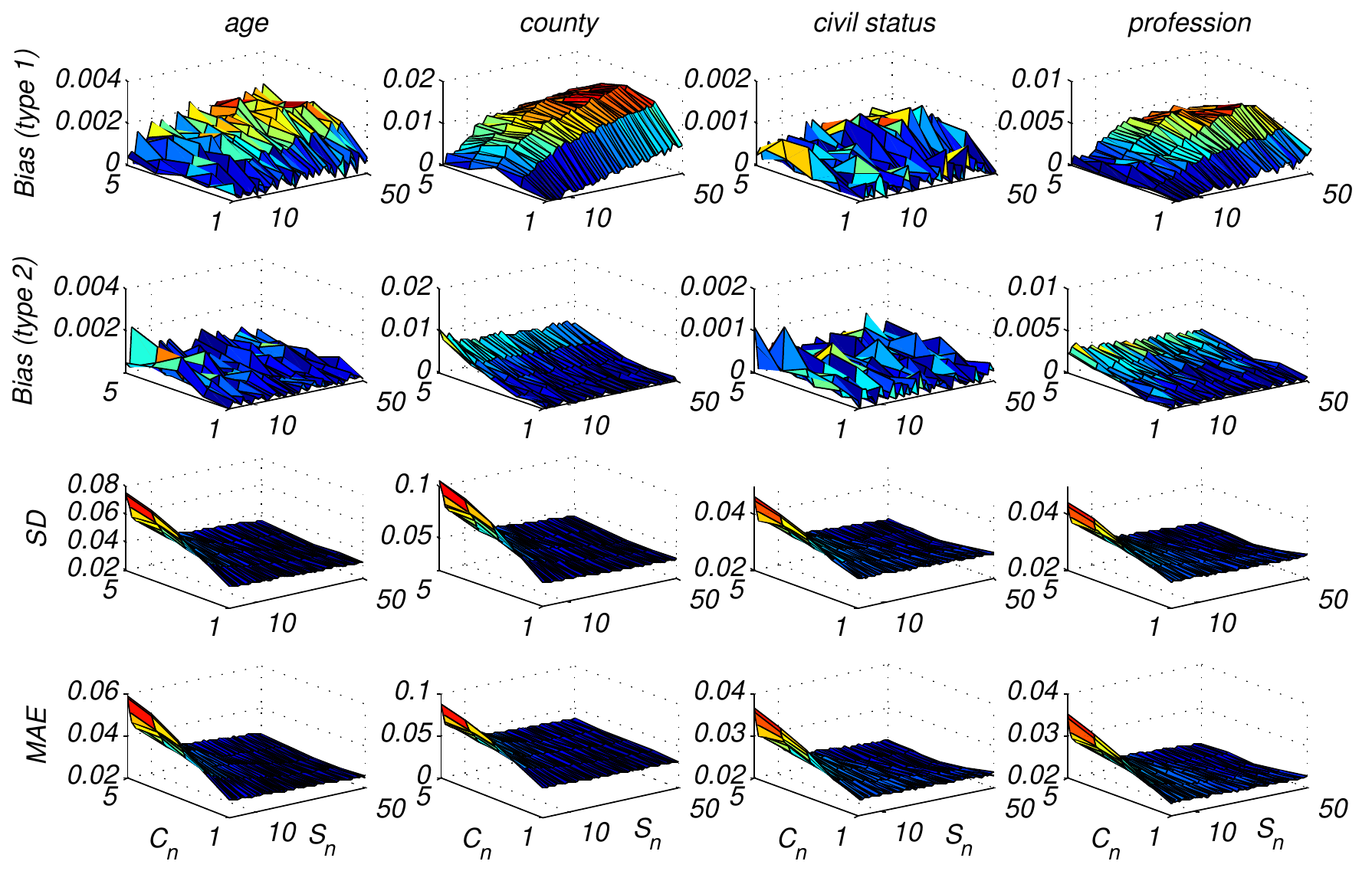}
\caption{Effects of varying the number of seeds and coupons in $G1$ when sample size was 500, with replacement. Simulation repeated 10,000 times for each combination. $C_n$ stands for the number of coupons and $S_n$ for the number of seeds.}\label{fig7}
\end{figure}

Determination of the number of coupons per participant and the number and characteristics of the seeds are among the first problems encountered by researchers when preparing an RDS study. To evaluate the effects of variations in these parameters, we increased the number of coupons and seeds, with the seed(s) being selected either with uniform probability (type 1) or with probability proportional to the nodes' degree (type 2). Results are shown in \autoref{fig7}. The difference between selection types for both SD and MAE were minute, and we therefore do not show SD and MAE for different selection types separately.

The first impression of this test is that an RDS study started by a type 2 selection seems to have a somewhat smaller bias than an RDS study started by a type 1 selection. The difference is slightly larger for \textit{age} and \textit{county}, which have larger homophilies. The number of seeds and coupons had small effects on the average \textit{RDS\uppercase\expandafter{\romannumeral2}} estimates, but had an obvious effect on the SD and MAE: both the SD and MAE increased when the samplings used more coupons. This is probably because a study with a small number of coupons needs longer chains to reach the same sample size. Longer chains, in turn, are more likely to break out of homogenous sub-networks and therefore become more representative of the overall population. Simulations on $G1_{add}$ and $G1_{rand}$ point to the same conclusions and are shown in \hyperlink{App-F}{Appendix F}. Real-life RDS studies cannot however use too few coupons or seeds as they will fail in generating long recruitment chains which are long enough.

\section{Conclusions}
\label{Conclusions}

Real social networks and the recruiting behavior of people in those networks can hardly meet all of the assumptions underlying the RDS estimators. Therefore it is crucial to know how much estimates are affected by deviations from these assumptions. This paper describes to the best of our knowledge, the first study simulating RDS studies within an actual hidden population and which in doing so can compare true population values to estimates obtained under various deviations from the \textit{RDS\uppercase\expandafter{\romannumeral2}} assumptions.

Our simulations show that when all the assumptions underlying the \textit{RDS\uppercase\expandafter{\romannumeral2}} estimator are fulfilled, results are excellent and asymptotically biased.

Further we show that the estimates adjusted by the eigenvector of eigenvalue 1 for the transition matrix are always unbiased for all groups and networks. This is further the reason why the \textit{RDS\uppercase\expandafter{\romannumeral2}} estimator is biased when the network is directed or when recruitment is a non-random selection from the participants' social networks. We should note that currently this eigenvector cannot be inferred from the empirical data in an RDS study where only the information of recruitment chain and personal network sizes are known. However, our findings show that given knowledge about this eigenvector it is possible with RDS to make unbiased estimates even on directed networks.

It turns out that when sample sizes were relatively small, sampling without replacement, which is used in practice, actually had a slightly lower standard deviation and mean absolute error than simulations with replacement.

The \textit{RDS\uppercase\expandafter{\romannumeral2}} estimator shows strong resistance to recruitment error when the probabilities that individuals will ignore contacts and reject invitations are independent of the individuals' characteristics. On the other hand, if these probabilities are dependent on the individuals' characteristics and if these characteristics are correlated with the outcome characteristic one wishes to estimate, the bias and MAE could become very large. As participants in RDS studies are rewarded for successful recruitments, rational and self-interested participants could be expected to ignore contacts who are considered less likely to accept invitations. Simulations show that such a combination of a group having a high probability of rejecting invitations and a high probability of being ignored can give rise to very large bias and error. We suggest that RDS studies should routinely compare participants reported network composition with actual recruitments to provide further empirical evidence on this issue from a wide variety of contexts.

Besides testing the violation of assumptions, we also analyzed some of the effects of network structure and homophily on the \textit{RDS\uppercase\expandafter{\romannumeral2}} estimator. The results are consistent with previous studies \citep{Volz2008,Gile2009,Goel2009}: networks which were sparse and had a skewed degree distribution had larger error and bias, and estimations of groups with small homophily performed better than estimations among groups with high homophily.

The deviations from the assumptions that we have simulated in this paper can be modeled in different ways, which affects the conclusions. We have opted for deviations that we consider relevant to RDS studies in different contexts, but the simulations still represent subjective choices and do not cover all situations relevant to real life RDS studies. Moreover, while we have tried to make results more generalizable by varying the properties of the original network, the network characteristics actually picked for simulations do not reflect all types of networks, which could impact on the interpretations of the results. For further studies it would be valuable to look at the effects of combinations of violations of the assumption in order to let the simulations better approximate reality.

\pdfbookmark[1]{Acknowledgement}{Acknowledgement}
\section*{Acknowledgement}
\label{Acknowlegement}
This work is funded in part by the Riksbankens jubileumsfond (dnr: P2008-0674). B.J.K. acknowledges the support from the Korea Science and Engineering Foundation (Grant No. R01-2007-000-20084-0), and X.L. would like to thank China Scholarship Council (Grant No. 2008611091).

\clearpage
\newpage
\pdfbookmark[1]{Appendix}{Appendix}
\section*{Appendix}\label{Appendix}
\pdfbookmark[2]{App-A Degree distribution for edge-added and edge randomized networks}{App-A}
\hypertarget{App-A}{}
\subsection*{App-A Degree distribution for edge-added and edge randomized networks}\label{APP-A}

\begin{figure}[htb]
\centering
\includegraphics[width=1\textwidth]{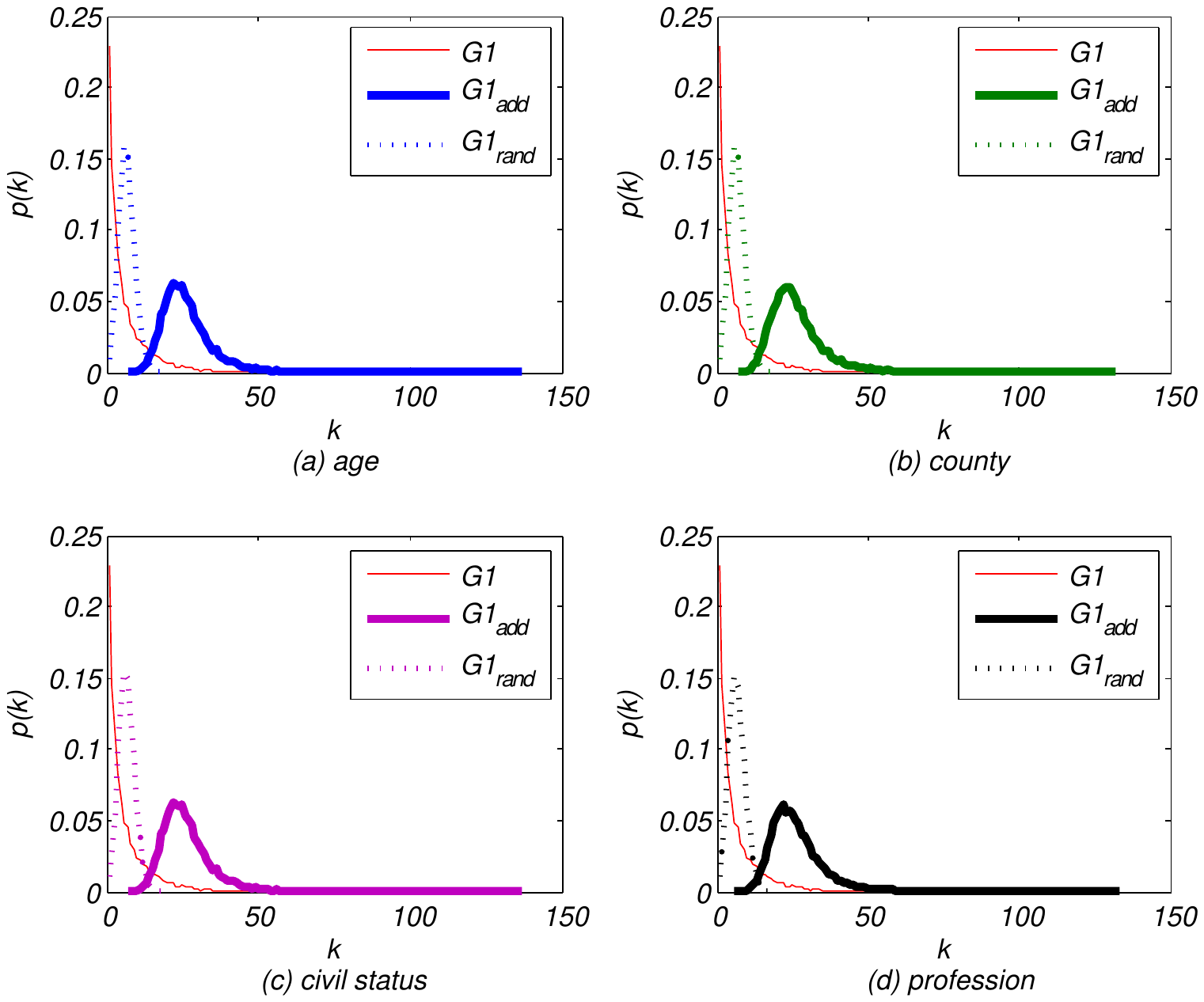}
\caption{Degree distribution for edge-added and edge-randomized networks. Red solid lines stand for the original undirected network ($G1$), bold lines for networks whose average degrees were increased by 20 ($G1_{add}$), and dashed lines for networks whose edges were randomly rewired ($G1_{rand}$)}\label{AppA}
\end{figure}

\clearpage
\newpage
\pdfbookmark[2]{App-B RDS on undirected networks: small sample sizes}{App-B}
\hypertarget{App-B}{}
\subsection*{App-B RDS on undirected networks: small sample sizes}\label{APP-B}

\begin{figure}[htb]
\centering
\includegraphics[width=1\textwidth]{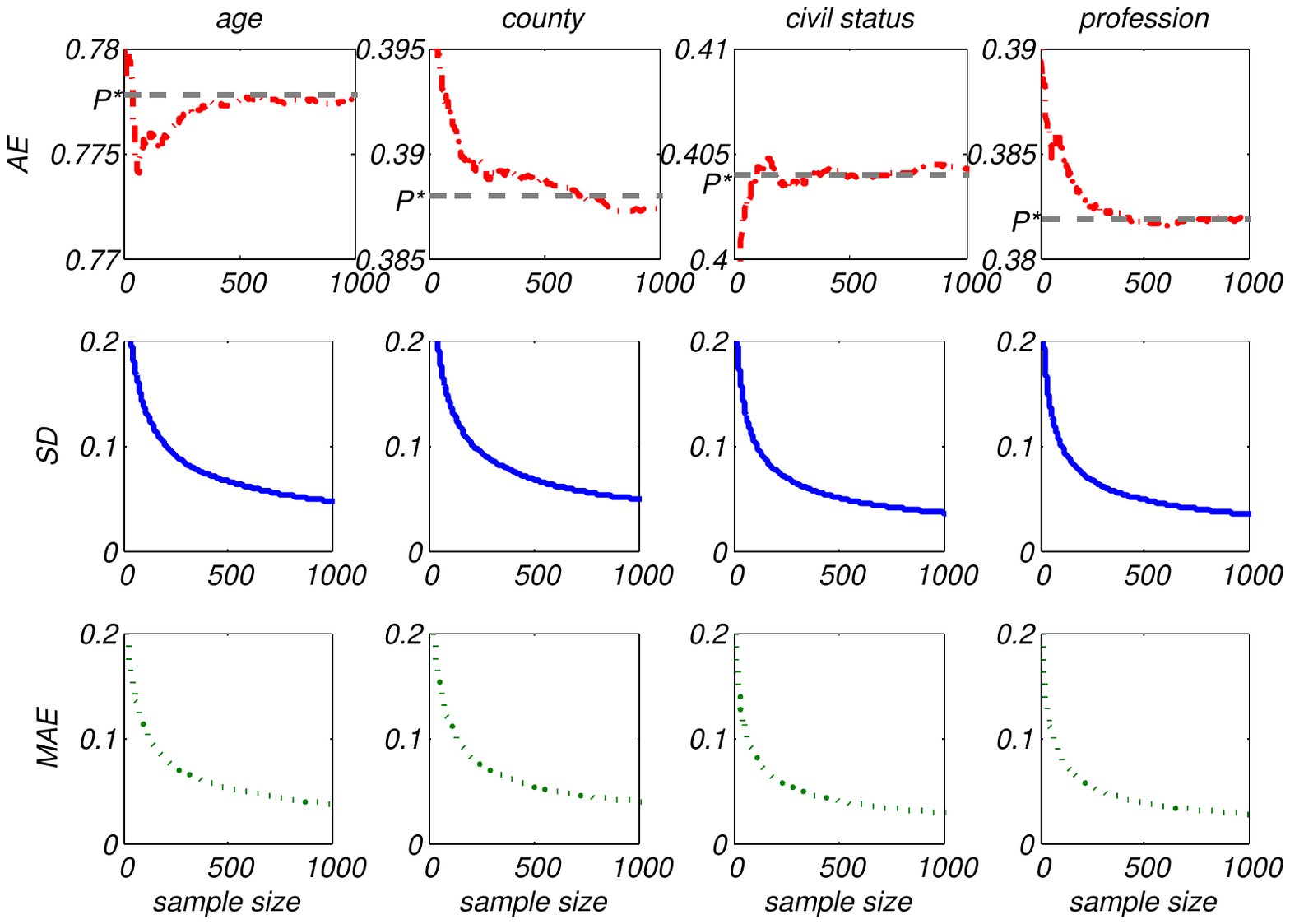}
\caption{\textit{RDS\uppercase\expandafter{\romannumeral2}} estimations on the undirected network ($G1$). The average estimates approach the true proportions very fast. When the sample size was 500 the Bias was only 0.0002, 0.0009, 0.00002, 0.0002 for \textit{age}, \textit{county}, \textit{civil status} and \textit{profession}, respectively.}\label{AppB}
\end{figure}

\clearpage
\newpage
\pdfbookmark[2]{App-C Sampling without replacement}{App-C}
\hypertarget{App-C}{}
\subsection*{App-C Sampling without replacement}\label{APP-C}

\begin{figure}[htb]
\centering
\includegraphics[width=1\textwidth]{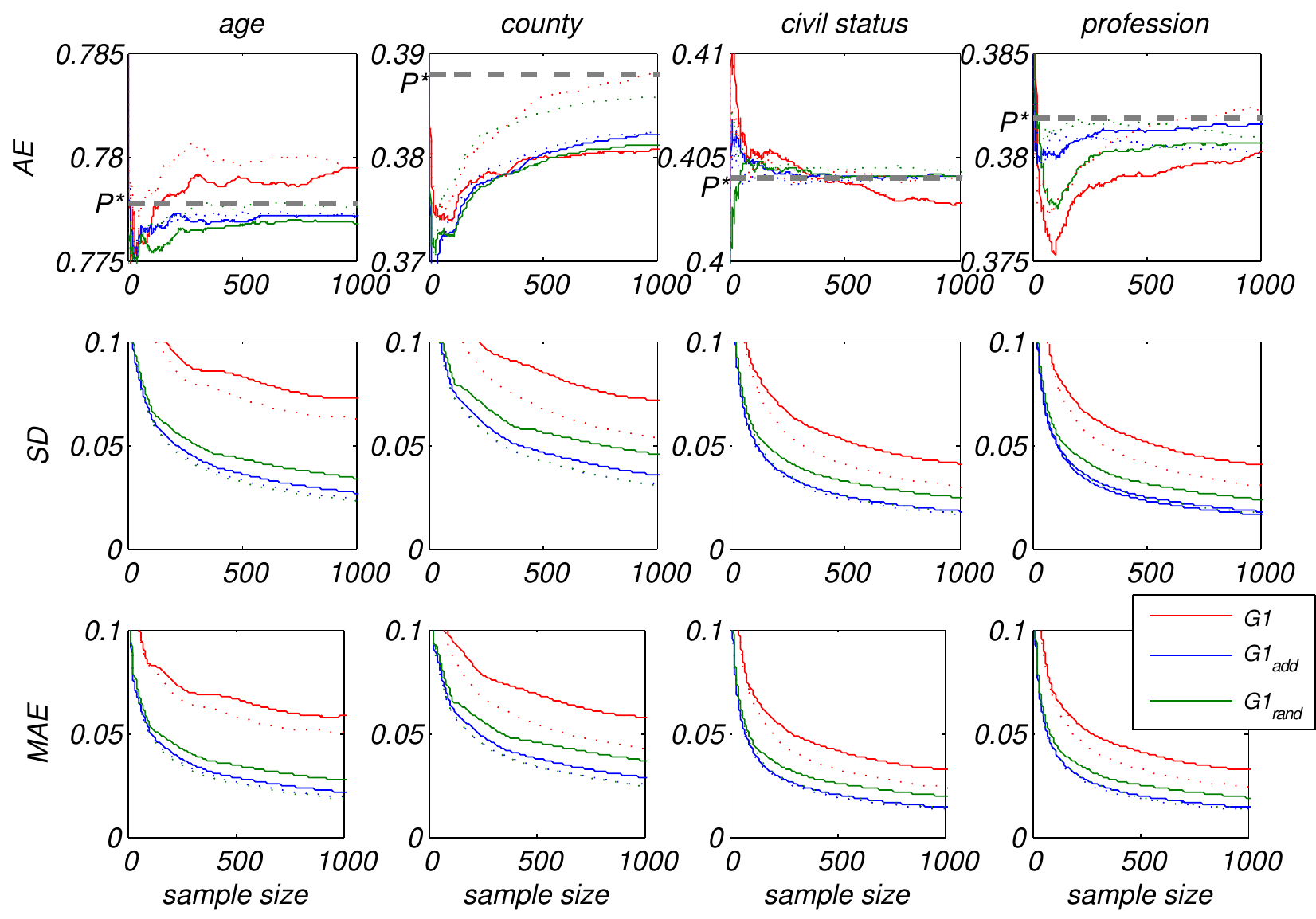}
\caption{Effects of network structure and replacement. Number of seeds=10, coupons=3. Seeds were randomly selected at the beginning of each simulation. Solid lines represent sampling with replacement and dashed lines represent sampling without replacement.}\label{AppC}
\end{figure}

\clearpage
\newpage
\pdfbookmark[2]{App-D Sampling with ignore and reject probabilities}{App-D}
\hypertarget{App-D}{}
\subsection*{App-D Sampling with ignore and reject probabilities}\label{APP-D}

All simulations were repeated 10,000 times for each combination of probabilities, number of seeds=10, coupon=3, with replacement. Seeds were randomly selected at the beginning of each simulation. Estimates were calculated when sample sizes reached 500.

\pdfbookmark[3]{App-D1 RDS with ignore and reject probabilities independent of the individuals' characteristics}{App-D1}
\hypertarget{App-D1}{}
\subsubsection*{App-D1 RDS with ignore and reject probabilities independent of the individuals' characteristics}\label{APP-D1}
\begin{figure}[htb]
\centering
\includegraphics[width=0.65\textwidth]{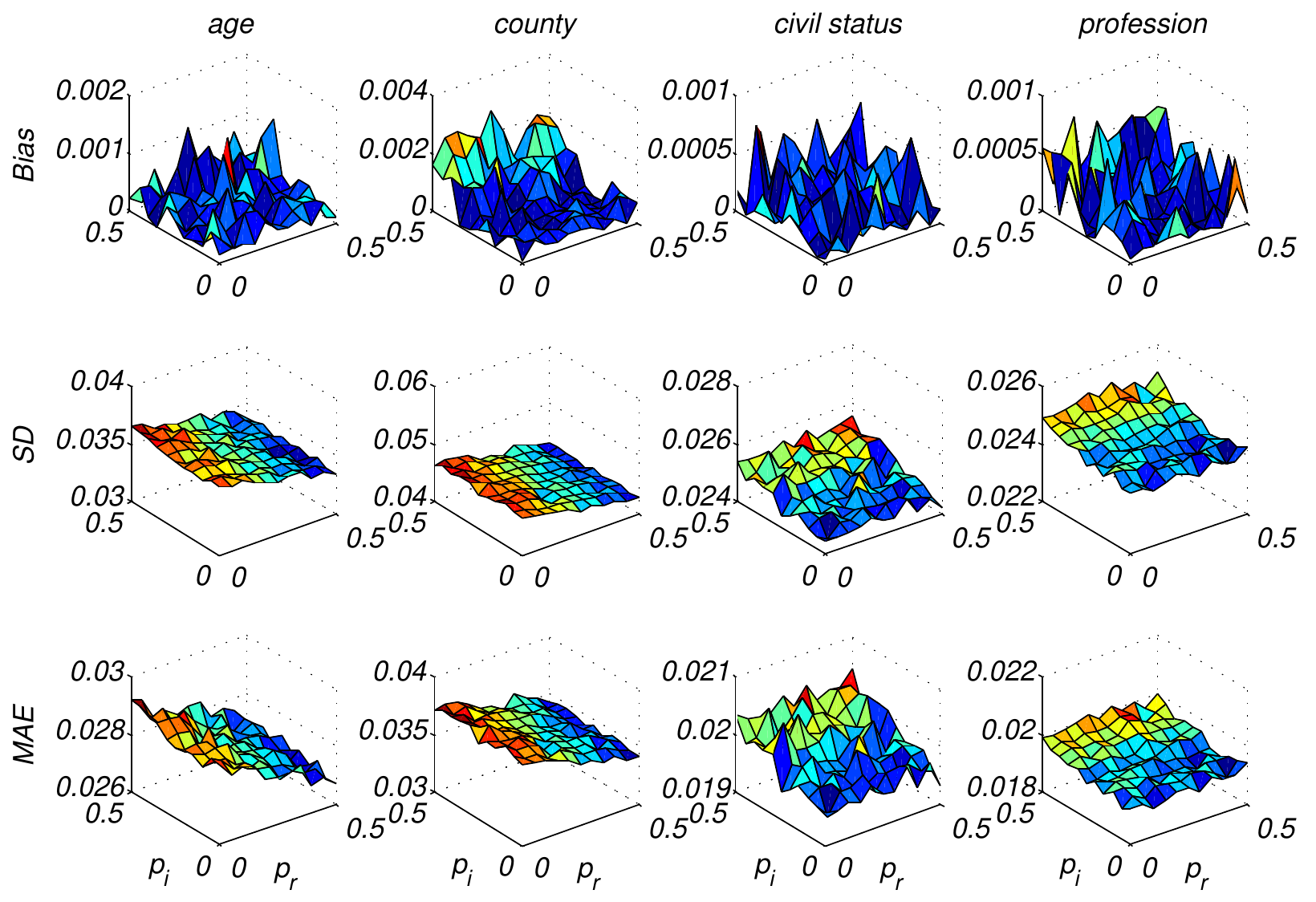}
\caption{Sampling with ignore and reject probabilities in the edge-added networks ($G1_{add}$).}\label{AppD_add}
\includegraphics[width=0.65\textwidth]{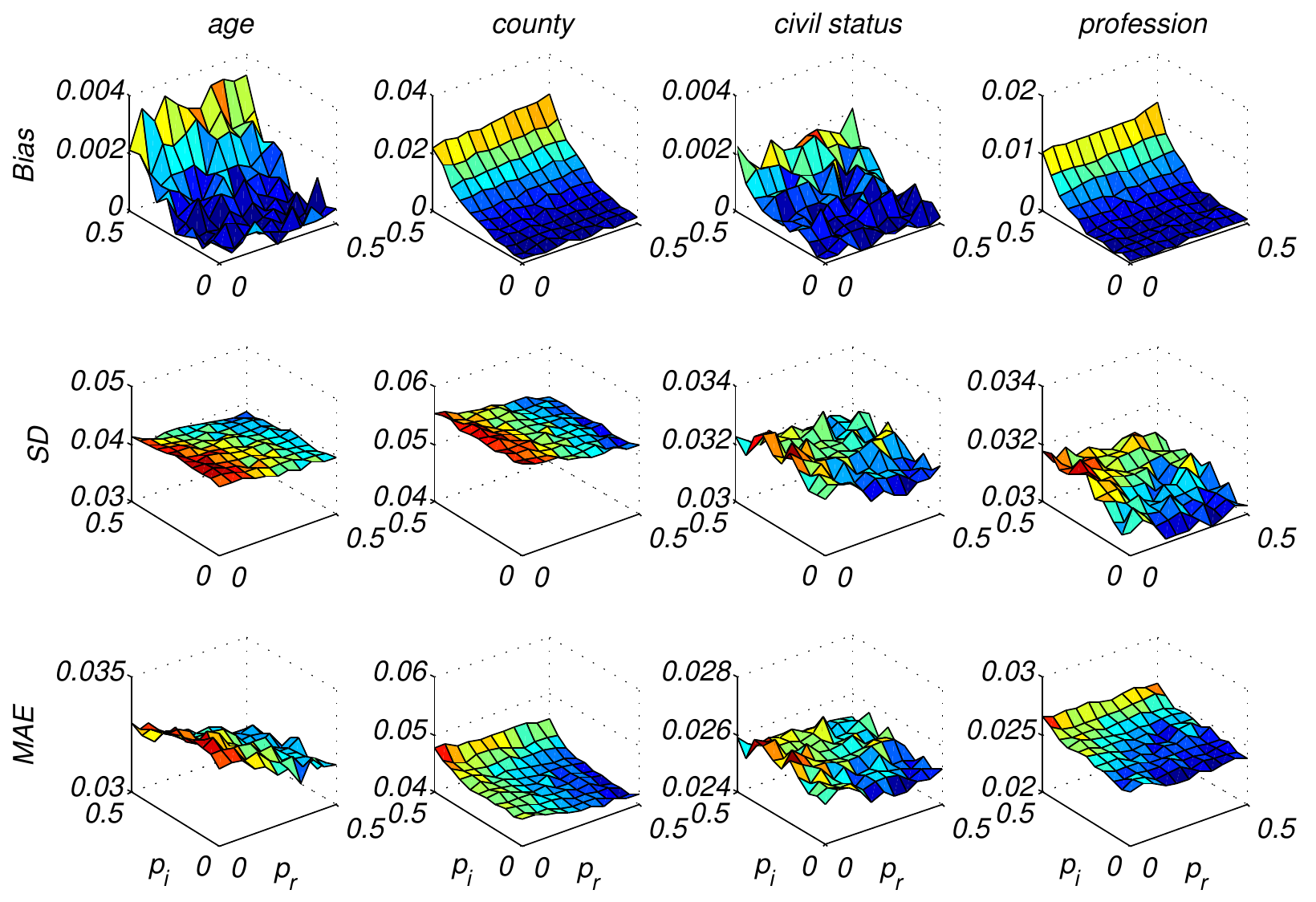}
\caption{Sampling with ignore and reject probabilities in the edge-randomized networks ($G1_{rand}$).}\label{AppD_rand}
\end{figure}

\clearpage
\newpage
\pdfbookmark[3]{App-D2 RDS with ignore and reject probabilities dependent on the individuals' characteristics}{App-D2}
\hypertarget{App-D2}{}
\subsubsection*{App-D2 RDS with ignore and reject probabilities dependent on the individuals' characteristics}\label{APP-D2}

\begin{figure}[htb]
\centering
\includegraphics[width=0.9\textwidth]{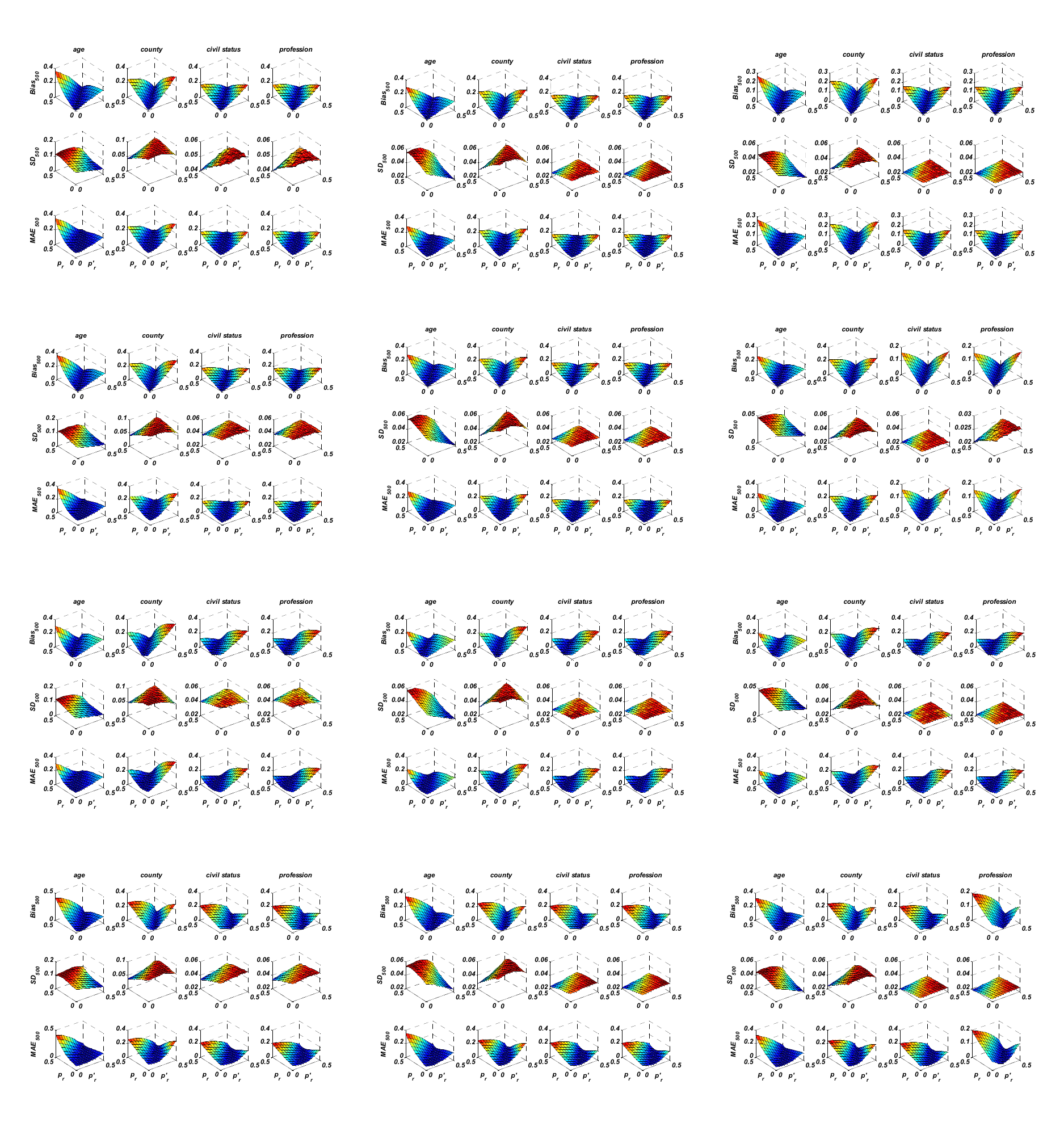}
\caption{RDS with ignore and reject probabilities. From top to bottom: $p_i=0$, $p'_i=0$; $p_i=0.2$, $p'_i=0.2$; $p_i=0.1$, $p'_i=0.3$; $p_i=0.3$, $p'_i=0.1$. (Left: $G1$, Middle: $G1_{rand}$, Right: $G1_{add}$)}
\end{figure}

\begin{figure}[htb]
\centering
\includegraphics[width=0.9\textwidth]{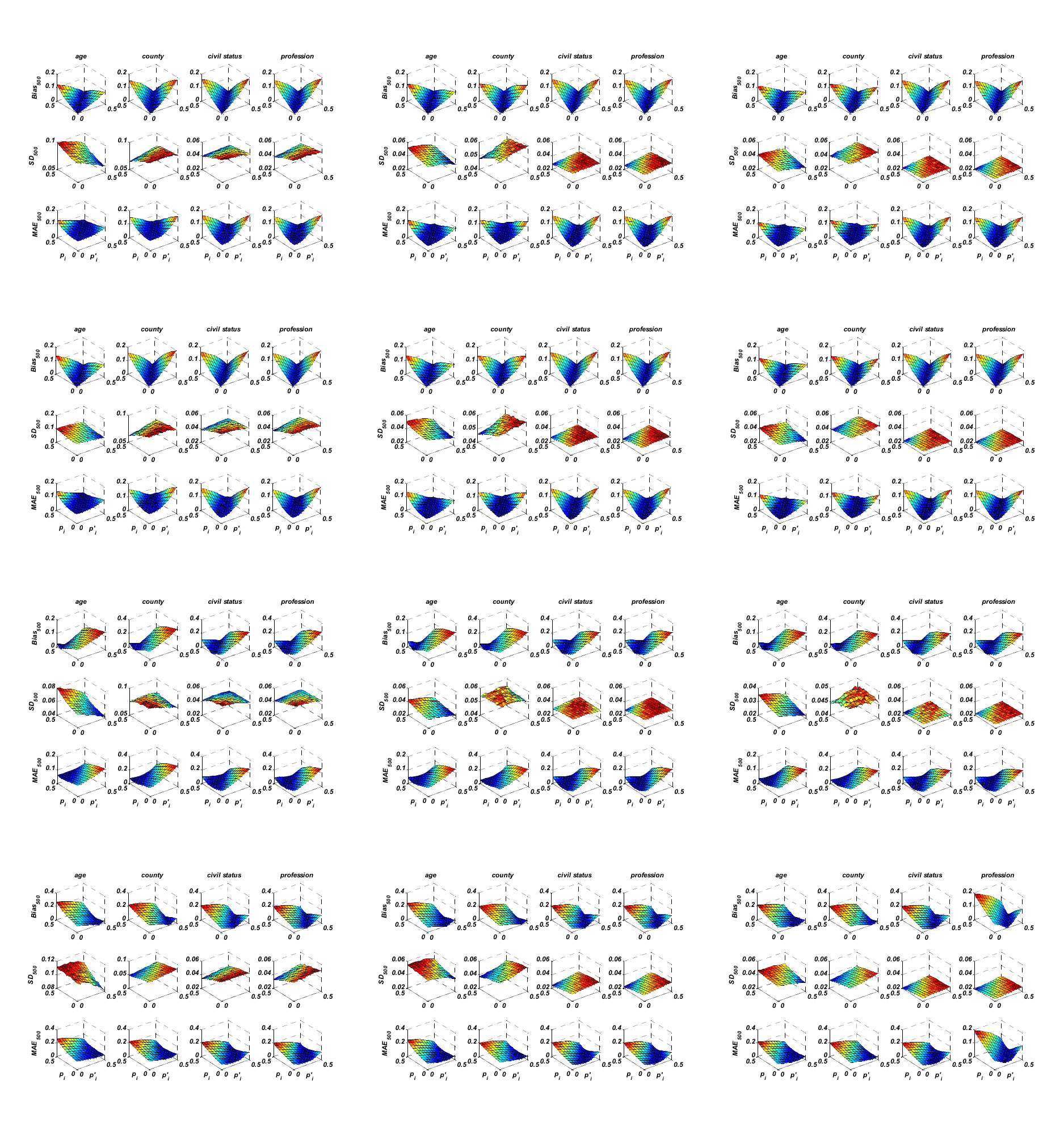}
\caption{RDS with ignore and reject probabilities. From top to bottom: $p_r=0$, $p'_r=0$; $p_r=0.2$, $p'_r=0.2$; $p_r=0.1$, $p'_r=0.3$; $p_r=0.3$, $p'_r=0.1$. (Left: $G1$, Middle: $G1_{rand}$, Right: $G1_{add}$)}
\end{figure}

\clearpage
\newpage
\pdfbookmark[2]{App-E Sampling with preferential recruitment}{App-E}
\hypertarget{App-E}{}
\subsection*{App-E Sampling with preferential recruitment}\label{APP-E}

\begin{figure}[htb]
\centering
\includegraphics[width=1\textwidth]{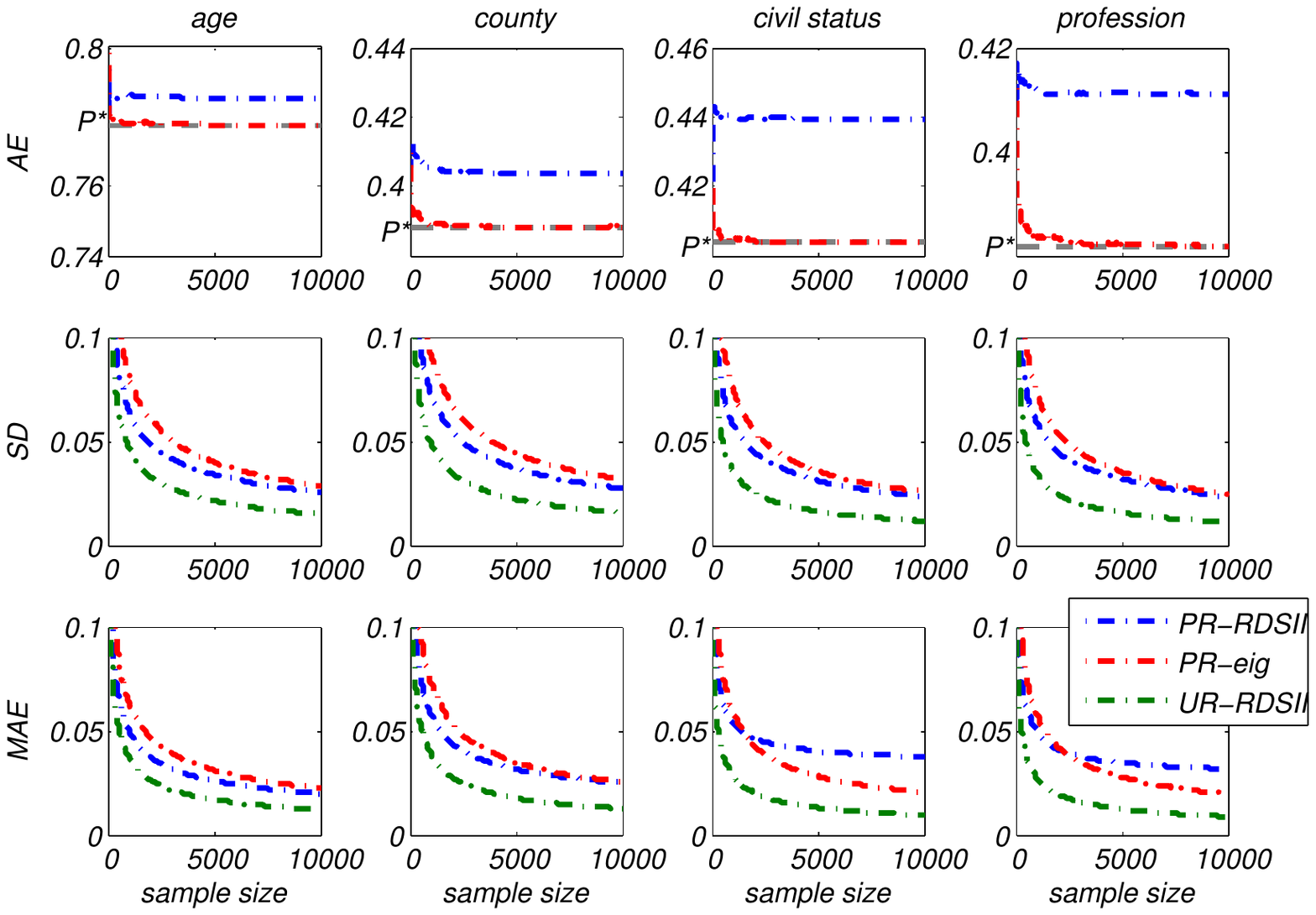}
\caption{RDS on $G1_{min}$ with preferential recruitment. Number of seeds=1, coupons=1, sampling with replacement. Seeds were randomly selected at the beginning of each simulation. The blue lines represent estimations by \textit{RDS\uppercase\expandafter{\romannumeral2}} and red lines represent estimations by eigenvector. Green lines are the \textit{RDS\uppercase\expandafter{\romannumeral2}} estimations for recruitment with uniform probability. Dashed gray lines indicate the true population values.}\label{AppE}
\end{figure}

\clearpage
\newpage
\pdfbookmark[2]{App-F Effect of seeds and coupons}{App-F}
\hypertarget{App-F}{}
\subsection*{App-F Effect of seeds and coupons}\label{APP-F}

\begin{figure}[htb]
\centering
\includegraphics[width=0.55\textwidth]{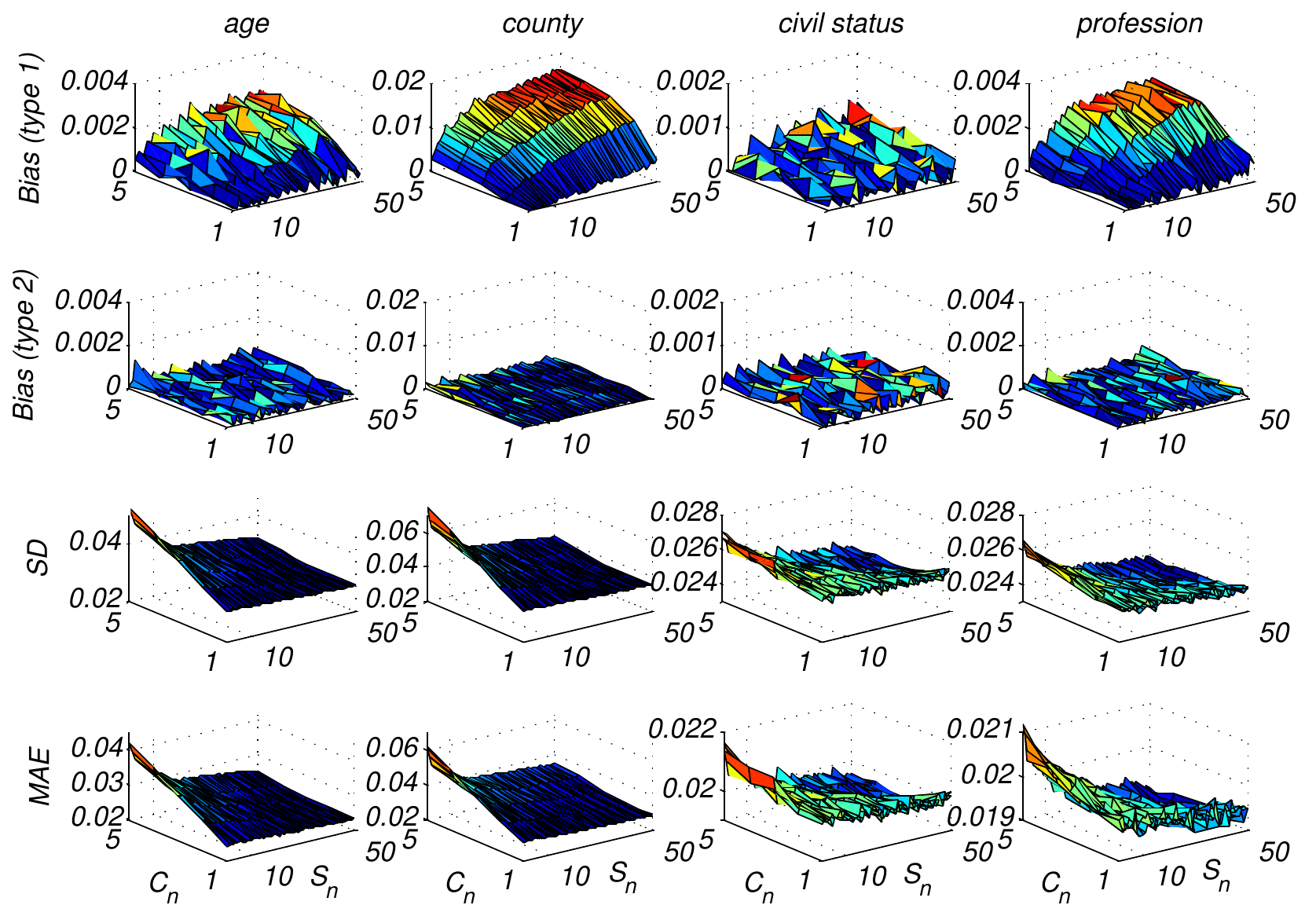}
\caption{Effects of varying the number of seeds and coupons in $G1_{add}$ when sample size was 500, with replacement. Simulation repeated 10,000 times for each combination. $C_n$ stands for the number of coupons and $S_n$ for the number of seeds.}\label{AppF_add}
\includegraphics[width=0.55\textwidth]{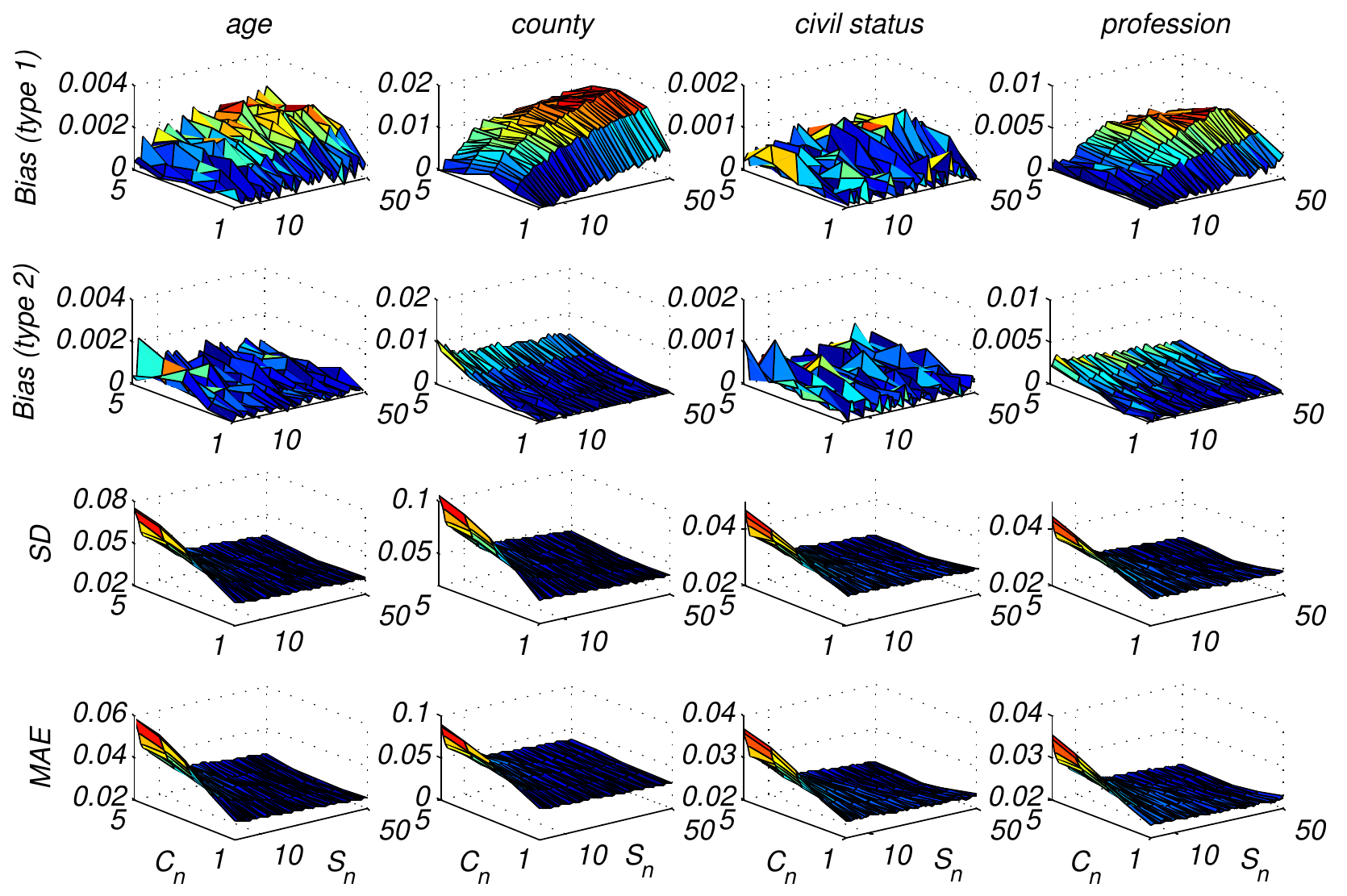}
\caption{Effects of varying the number of seeds and coupons in $G1_{rand}$ when sample size was 500, with replacement. Simulation repeated 10,000 times for each combination. $C_n$ stands for the number of coupons and $S_n$ for the number of seeds.}\label{AppF_rand}
\end{figure}

\clearpage
\newpage
\pdfbookmark[1]{References}{References}
\section*{References}
\bibliography{RDS_Sensitivity_arXiv}
\bibliographystyle{elsarticle-harv}
\end{document}